\documentclass[aps,a4paper,10pt,twocolumn,nofootinbib]{revtex4} 
\usepackage{datetime}
\usepackage{amsmath}
\usepackage{amsfonts}
\usepackage{amsfonts}
\usepackage{mathrsfs}
\usepackage[mathscr]{euscript}
\usepackage[dvips]{graphicx}
\usepackage{fancyhdr}
\usepackage{colordvi}
\usepackage{epsfig}
\usepackage{color}
\usepackage{bm}
\usepackage{soul}

\def\overstrike#1#2{{\setbox0\hbox{$#2$}\hbox to \wd0{\hss
    $#1$\hss}\kern-\wd0\box0}}
\newcommand{\convol}{\star}

\def\doiURL#1{doi:\href{http://dx.doi.org/#1}{#1}}
\def\axURL#1{arxiv:\href{http://arxiv.org/abs/#1}{#1}}

\newdateformat{yymmdddate}{\THEYEAR/\twodigit{\THEMONTH}/\twodigit{\THEDAY}}

\numberwithin{equation}{section}

\begin{document}
\title[Acoustic waves: propagated in time or space?]
      {Acoustic waves: should they be propagated forward in time, or forward in space?}
\author{Paul \surname{Kinsler}}
\email{Dr.Paul.Kinsler@physics.org}
\affiliation{
  Blackett Laboratory, Imperial College London,
  Prince Consort Road,
  London SW7 2AZ, 
  United Kingdom.
}

\lhead{\includegraphics[height=5mm,angle=0]{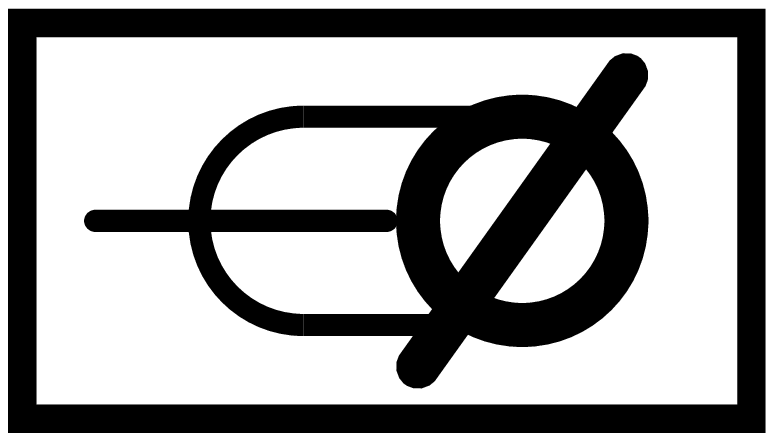}~~FBACOU}
\chead{Acoustic waves: propagated in time or space?}
\rhead{
\href{mailto:Dr.Paul.Kinsler@physics.org}{Dr.Paul.Kinsler@physics.org}\\
\href{http://www.kinsler.org/physics/}{http://www.kinsler.org/physics/}
}

\begin{abstract} 

The evolution of acoustic waves can be evaluated in two ways:
 either as a temporal, or a spatial propagation.
Propagating in space provides the considerable advantage 
 of being able to handle dispersion and propagation across interfaces
 with remarkable efficiency; 
 but propagating in time is more physical
 and gives correctly behaved reflections and scattering without effort.
Which should be chosen in a given situation, 
 and what compromises might have to be made?
Here the natural behaviors of each choice of propagation
 are compared and contrasted for 
 an ordinary second order wave equation, 
 the time-dependent diffusion wave equation,
 an elastic rod wave equation, 
 and the Stokes'/ van Wijngaarden's equations, 
 each case illuminating a characteristic feature of the technique.
Either choice of propagation axis enables
 a partitioning the wave equation that 
 gives rise to a directional factorization
 based on a natural ``reference'' dispersion relation.
The resulting exact coupled bidirectional equations
 then reduce
 to a single unidirectional first-order wave equation
 using a simple ``slow evolution'' assumption
 that minimizes effect of subsequent approximations, 
 while allowing a direct
 term-to-term comparison between exact and approximate theories.

\end{abstract}






  


\date{\today}
\maketitle
\thispagestyle{fancy}

%

%
\section{Introduction}\label{S-intro}

An important category of acoustic wave models 
 consists of those based on second order wave equations.
Because they include second order derivatives 
 in both time
 and space, 
 they naturally lend themselves
 to rearrangements designed to focus either 
 on the temporal or the spatial behaviour.
In practice, 
 most acoustic wave models have more than just 
 this pair of second order terms, 
 and these extra contributions may (or may not)
 be suited to a preference for either temporal or spatial analysis.
But despite the wide range of different acoustic wave models,
 a subset of cases is sufficient 
 to cover most important modifications.
Here we choose 
 the time-dependent diffusion equation
 (TDDE) \cite{Arfken-MMfPhy}, 
 a model for waves in an elastic cylindrical rod \cite{Porubov-ANLSWS}, 
 and acoustic waves 
 as described by a generalized Stokes' equation \cite{Stokes-1845tcps}
 which allows for bubbles  \cite{Wijngaarden-1972arfm}.
For these three situations, 
 harmonic solutions are well known, 
 but transient solutions are harder to find.
Recently, 
 the impulse-response and causal properties 
 of some of these wave equations have been analyzed by 
 Buckingham \cite{Buckingham-2008jasa},
 and the (perhaps surprising) ``non-causal'' claims
 made therein are briefly addressed.

An advantage of second order wave equations is that 
 once a propagation type is selected --
 whether into the future (i.e. along time), 
 or along an appropriate spatial axis --
 they can be decomposed into pairs of 
 explicitly directional and \emph{first order} acoustic wave equations.
The resulting coupled first order equations are not only often 
 easier and faster to solve than the 
 starting point of a second order wave equation,
 they are also well suited to the case of 
 unidirectional traveling waves.
However the unidirectional approximation is not demanded, 
 and the formulation means that any additional approximations
 tend to be less restrictive
 \cite{Kinsler-2010pra-fchhg}.
For the purposes of this article, 
 however, 
 these desirable features are not the core message.
Instead, 
 by comparing the results of factorizations
 aimed at generating \emph{temporally} propagated wave equations,
 with those that generate \emph{spatially} propagated ones,
 comparisons and contrasts can be made
 between the natural ``reference'' behaviors
 present in each decomposition. 
We will see that 
 some terms in a wave equation 
 will lend themselves to inclusion in one reference behaviour
 but not the other.
These results then inform us as to how we might choose to trade off the 
 practical efficiency of a spatially propagated picture
 against the physically accurate temporally propagated one.

Section \ref{S-decompose} provides an overview of the 
 factorization process for both spatial and temporal decompositions, 
 using the time domain diffusion equation as an example.
Factorization is then used to analyze 
 the TDDE in section \ref{S-TDDE},
 acoustic waves in 
 an elastic cylindrical rod in section \ref{S-ERod}, 
 and the Stokes' equation and its bubbly generalization
 in section \ref{S-VWE}.
This variety allows a discussion of both common and contrasting features
 of these directional decompositions.
The conclusions are given in section \ref{S-conclude}.

%


%
\section{Directional decompositions}\label{S-decompose}

To compare the physical representations of waves
 in temporal or spatially propagated pictures, 
 we use a factorization technique.
This makes use of the concept
 of ``reference'' or ``underlying'' wave evolution
 \cite{Kinsler-2010pra-fchhg,Kinsler-2010pra-dblnlGpm}.
If these reference behaviors are a close match to the 
 actual wave evolution, 
 subsequent approximations will be less stringent --
 in particular if we choose to make a unidirectional approximation.
Although in some systems an exact match is possible, 
 this is rarely the case if the waves depart from 
 some idealized linear behaviour.
Nevertheless, 
 the power of these directional decompositions
 is in considering such variation from these exactly solvable cases,
 situations which often require approximation or numerical integration.
To ``factorize'', 
 start by selecting a propagation axis -- 
 either time, 
 or along some spatial trajectory --
 after which the total wave can be decomposed into directional components
 that evolve forwards or backwards 
 perpendicular to that propagation.

Throughout this article the context and/or arguments 
 determine which domains ($\Vec{r}$ or $\Vec{k}$, $t$ or $\omega$)
 a given instance of some relevant function covers; 
 in addition some symbols 
 (e.g. $Q$, $\Omega$, $\kappa$, $c$, $c_\Omega$, $c_\kappa$, etc)
 are reused independently of each other 
 (for different wave equation models) 
 without distinguishing subscripts -- 
 this is in order to avoid cluttering the notation.
Further, 
 each wave equation given has, 
 on the right hand side (RHS), 
 a term $Q$ which represents some general source term.
Typically \cite{Buckingham-2008jasa,Grulier-PTPR-2009jasa}
 $Q$ is an impulse designed to elicit the primitive response
 of the system
 (i.e. $Q = Q \delta(t)\delta(x))$,
 but here it is allowed to be any kind of perturbation, 
 (non)linear modification, 
 or driving term we desire.
For example, 
 in the simple wave equations considered in this section, 
 a density $\rho(\Vec{r})$ could be added,
 setting $Q = (1/\rho) \nabla \rho \cdot \nabla g$,
 and so match a wave equation used for ultrasound propagation
 \cite{Mari-BMUC-2009jasa,Jensen-1991jasa}.
Alternatively, 
 adding a loss term to $Q$ with the form $\eta \partial_t g$
 would give us the 
 time-dependent diffusion equation (TDDE) \cite{Arfken-MMfPhy},
 which appears in a variety of contexts in physics, 
 including acoustic waves in plasmas 
 or the interstitial gas filling a porous,
 statistically isotropic, 
 perfectly rigid solid \cite{MorseIngard-Acoustics};
  it also models electromagnetic wave propagation through
     conductive media and is known as the telegrapher's equation.

In this section, 
 an ordinary second order wave equation for 
 some appropriate property $g(\Vec{r},t)$ will be used
 as a test bed
 on which to demonstrate the factorization process.
It contains both a temporal response function $p(t)$
 and a spatial one $s(\Vec{r})$, 
 both have a non-local character
 which is allowed for using a convolution (``$\convol$''), 
 with $a(u) \convol b(u) \equiv \int a(u')b(u-u')du'$.
Whilst some materials, 
 in some regimes, 
 do indeed have non-local spatial properties, 
 the role of $s(\Vec{r})$ here is to ensure that
 the wave equation includes some nontrivial spatial properties.
In many situations 
 spatial structure would be introduced using $p(\Vec{r},t)$
 instead of just $p(t)$,
 while $s(\Vec{r})$ and its convolution would be absent; 
 however here that would introduce complications beyond the 
 scope of this example.
The wave equation is
~
\begin{align}
  c^2
  \nabla^2
    s(\Vec{r}) \convol
    g(\Vec{r},t)
 -
  \partial_t^2
    p(t) \convol
    g(\Vec{r},t)
&=
 c^2
 Q(x,t)
,
\label{eqn-simple-xt}
\end{align}
 where
 a non-interacting wave travels with speed $c$. 
Upon Fourier transforming into the $k,\omega$ domain,
 where $d/dt \equiv \partial_t \leftrightarrow -\imath\omega$
 and $\nabla \leftrightarrow +\imath \Vec{k}$, 
 with $k^2 = \Vec{k} \cdot \Vec{k}$,
 the result is
~
\begin{align}
  c^2
  k^2 s(\Vec{k})
    g(k,\omega)
 -
  \omega^2 p(\omega)
    g(\Vec{k},\omega)
&=
 -c^2 Q(\Vec{k},\omega)
,
\label{eqn-simple-kw}
\end{align}
 where $s(\Vec{k})$ is the spatial Fourier transform of $s(\Vec{r})$, 
 and $p(\omega)$ the temporal Fourier transform of $p(t)$.

In both styles of derivation that follow,
 the steps taken are a good mathematical match
 to typical spatial propagation procedures
 \cite{Ferrando-ZCBM-2005pre,Kinsler-2010pra-fchhg};
 but the physical meaning alters 
 with the interchange of the roles of time and space.
Factorization methods have a long history
 \cite{Infeld-H-1951rmp}, 
 but their adaption, application, and adoption to wave propagation
 of the type and context proposed here has been lacking
 until more recently
 \cite{Billger-F-2003wm,Peng-P-2004jasa,Jonsson-N-2010wm}.

Since $Q$ can contain any sort of behaviour, 
 eqn. \eqref{eqn-simple-kw}
 is already useful:
 e.g. factorization could easily be applied to 
 the nonlinear propagation 
 addressed by Pinton et al. \cite{Pinton-CGT-2010jasa}.
In that case
 their initial eq. (3) can be matched to eqn. \eqref{eqn-simple-xt} above
 by restricting to the $x$-axis, 
 using $c^2 s(x)=\mu \delta(x)$, 
 $p(t)=\rho \delta(t)$, 
 and setting $Q$ to match their RHS nonlinear term.
Using directional decomposition, 
 a first order wave equation 
 simpler than e.g. eq.(10) in Pinton et al. 
 can be obtained rapidly with fewer and less restrictive approximations.
Of course, 
 some acoustic wave equations reduced down to 
 apply to a single wave property 
 will not have the second order derivatives
 needed for this factorization scheme.
However, 
 such wave equations are already extensively approximated, 
 so there may be scope for factorizing related equations which
 do contain them.

%
\subsection{Temporal propagation, spatial decomposition}\label{S-decompose-time}

\begin{figure}
 \includegraphics[width=0.80\columnwidth,angle=0]{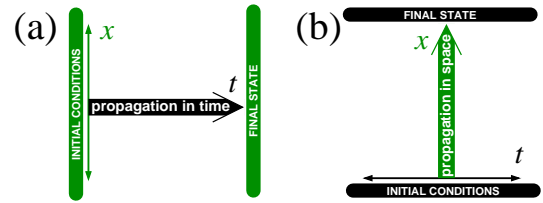}
\caption{
For temporal propagation (left), 
 initial conditions cover all space
 at an initial time $t_i$; 
 the final state at $t_f$ also covers all space.
For spatial propagation (right), 
 what are normally considered to be \emph{boundary} conditions
 take up the role of the 
 initial conditions; 
 at both the starting location $x_i$
 and the final location of $x_f$, 
 they specify the wave's behaviour over all time.
}
\label{F-diagram-propagator}
\end{figure}

The most physically motivated factorization is to 
 choose to \emph{propagate forward in time}, 
 and decompose the system behaviour (waves) into directional components
 that then evolve either forward or backward in space, 
 as shown in Fig. \ref{F-diagram-propagator}.
This is useful for analyzing situations 
 where signals need to be separated from reflections, 
 but requires an explicit modeling of the medium's time-response, 
 perhaps involving convolutions
 or auxiliary differential equations.

To obtain such a temporally propagated, 
 spatially decomposed representation,
 choose a suitable reference frequency $\Omega$.
This is allowed to have a wavevector ($k$) dependence, 
 but should not have a frequency ($\omega$) dependence.
Rewriting eqn. \eqref{eqn-simple-kw} for temporal propagation, 
~
\begin{align}
  \bar{\Omega}(\Vec{k},\omega)^2
    g(\Vec{k},\omega)
 -
  \omega^2
    g(\Vec{k},\omega)
&=
 -
  \frac{c^2}{p(\omega)} 
  Q(\Vec{k},\omega)
,
\label{eqn-simple-t-kw-first}
\end{align}
 with $\bar{\Omega}(\Vec{k},\omega)^2 = c^2 k^2 s(\Vec{k}) / p(\omega)$.
Unfortunately, 
 the frequency dependence of $\bar{\Omega}$ 
 means that it is not a good reference on which
 to decompose the propagation.
Thus we must either ignore the material's frequency response $p(\omega)$, 
 treat it as a correction by incorporating it into $Q$, 
 or convert its frequency response into a wavevector response $p(k)$
 by approximation\footnote{For example, 
 if restricted to waves near a frequency $\omega_1$
 in a narrow bandwidth $\Delta \omega \ll \omega_1$, 
 we can write $\omega \simeq c' k$, 
 where $c' = \omega_1/k$, 
 and replace $p(\omega)$ with $s(k)$ --
 i.e. convert a temporal dispersion 
 into an approximate spatial dispersion. 
 This
  is what was assumed in \cite{Carter-1995pra}.}.

If there is no temporal response, 
 then $p(t) \equiv p_0 \delta(t)$ and $p(\omega)=p_0$, 
 and the wave equation becomes 
~
\begin{align}
  \Omega(k)^2
    g(\Vec{k},\omega)
 -
  \omega^2
    g(\Vec{k},\omega)
&=
 -c^2 Q(\Vec{k},\omega) / p_0
,
\label{eqn-simple-t-kw}
\end{align}
 with $\Omega(\Vec{k})^2 = c^2 k^2 s(\Vec{k})^2/ p_0$.
Eqn. \eqref{eqn-simple-t-kw} can be rearranged
 to give an expression for $g(\Vec{k},\omega)$ directly, 
 i.e.
~
\begin{align}
   g
&=
  \frac{c^2}
       {\omega^2-\Omega^2}
  \frac{Q}{p_0} 
&=
 -
  \frac{1}
       {2 \Omega}
  \left[
    \frac{1}{\omega+\Omega}
   -
    \frac{1}{\omega-\Omega}
  \right]
  \frac{c^2 Q}{p_0} 
,
\label{eqn-Simple-kw-t-g}
\end{align}

Now decompose $g$ into wave components $g_{+}$ and $g_{-}$
 that evolve either forward or backward in space, 
 with $g=g_{+} + g_{-}$.
Then split eqn. \eqref{eqn-Simple-kw-t-g}
 into two coupled first order wave equations
 for $g_{\pm}(\Vec{k},t)$, 
 i.e.
~
\begin{align}
  g_{\pm}
&=
 \pm
  \frac{1}
       {2 \Omega}
    \frac{1}{\omega\mp\Omega}
  \frac{c^2 Q}{p_0} 
\\
  \left[
    \omega\mp\Omega
  \right]    
  g_{\pm}
&=
 \pm
  \frac{1}
       {2 \Omega}
  \frac{c^2 Q}{p_0} 
,
\label{eqn-Simple-kw-w-pre-gpm}
\end{align}
leading to either frequency or time domain forms
~
\begin{align}
  \omega
  g_{\pm}(\Vec{k},\omega)
&=
 \pm
  \Omega(k)
  g_{\pm}(\Vec{k},\omega)
 \pm
  \frac{1}
       {2 \Omega(\Vec{k})}
  \frac{c^2 Q(k,\omega)}{p_0}
,
\label{eqn-Simple-kw-w-gpm}
\\
  \partial_t
  g_{\pm}(\Vec{k},t)
&=
 \mp
  \imath
  \Omega(\Vec{k})
  g_{\pm}(\Vec{k},t)
 \mp
  \frac{\imath}
       {2 \Omega(\Vec{k})}
  \frac{c^2 Q(\Vec{k},t)}{p_0}
.
\label{eqn-Simple-kw-t-gpm}
\end{align}
This last equation tells us how $g_{\pm}(\Vec{k},t)$ evolves
[or if appropriately transformed, $g_{\pm}(\Vec{r},t)$]
 as we propagate in time.
The original second order equation
 can be recovered\cite{Ferrando-ZCBM-2005pre}
 by substituting one of these into the other.
Note that retaining a frequency dependence for $\Omega$
 would have given rise to time derivatives on the RHS
 of eqn. \eqref{eqn-Simple-kw-t-gpm}.
In addition to mathematical complications, 
 these would also disrupt
 the otherwise straightforward integration of $g_{\pm}$ in time.

If 
 $Q$ is small, 
 i.e. 
 $| c^2 Q / p_0 | \ll 2 | \Omega^2 g_{\pm} |$,
 then the wave evolves slowly as it propagates.
This ``slow evolution'' permits a 
 temporal version 
 of the unidirectional approximation 
 \cite{Kinsler-2010pra-fchhg,Kinsler-2007josab}
 to be made\footnote{However, 
  when $Q$ is a simple source term, 
  it merely acts to drive both forward and backward waves equally. 
  The two waves are then uncoupled, 
  and no $g_{+}$ dynamics are affected by the neglect of $g_{-}$, 
  and vice versa.
 If $Q$ has any dependence on $g$, 
  e.g. if it were a loss or nonlinear term, 
  then this would not be the case.}, 
 setting $g_{-}=0$.
The forward wave $g_{+}(\Vec{k},t)$ then follows
~
\begin{align}
  \partial_t
  g_{+}(\Vec{k},t)
&=
 -
  \imath
  \Omega(\Vec{k})
  g_{+}(\Vec{k},t)
 -
  \frac{\imath}
       {2 \Omega(\Vec{k})}
  \frac{c^2 Q(\Vec{k},t)}{p_0}
.
\label{eqn-Simple-kw-t-gp}
\end{align}
Note that if $Q$ is an impulse such as a delta function, 
 the slow evolution condition will not hold at that point.
If $\Omega^2 < 0$, 
 then $g$ or $g_{\pm}$ do not oscillate as time passes,
 but instead \emph{decay}.

%

%
\subsection{Spatial propagation, temporal decomposition}\label{S-decompose-space}

We could also choose to 
 \emph{propagate forward in space}, 
 and then decompose the system behaviour (waves) into components 
 that evolve either forward or backward in time, 
 as shown in fig. \ref{F-diagram-propagator}.
Although spatial propagation can seem non-intuitive, 
 and requires careful handling of reflections or scattering,
 spatial propagation schemes are nevertheless popular, 
 as in e.g. the KZK equation
 \cite{Zabolotskaya-K-1969spa,Kuznetso-1971spa}.
Their primary advantage is that 
 the full (past and future)
 time history of a wave is available at whatever point in space
 the system has reached; 
 in addition
 transitions across interfaces are also easy to handle.
Thus medium parameters such as speed or wavevector
 can be defined as functions of frequency, 
 so that material dispersion can be easily handled 
 using pseudospectral methods \cite{Fornberg-PSmethods,Tyrrell-KN-2005jmo}, 
 even enabling numerical investigations
 into exotic phenomena
 by means of artificial dispersions \cite{Kinsler-RTN-2007pre}.
    The angular spectrum approach 
       \cite{Stepanishen-B-1982jasa,Williams-M-1982jasa}
       is another pseudospectral method, 
       widely used in underwater and biomedical acoustics.

To obtain a spatially propagated, 
 temporally decomposed representation, 
 we choose $x$ as the propagation axis, 
 so that $y, z$ are the transverse spatial; dimensions, 
 with $\nabla_T^2 = \partial_y^2 + \partial_z^2$ 
 and transverse wavevectors $k_y$ and $k_z$ with $k_T^2 = k_y^2 + k_z^2$.
Next, 
 we need a suitable reference wavevector $\kappa$, 
 which is allowed to have a frequency ($\omega$) dependence, 
 but should not have a wavevector ($k$) dependence.
Rewriting eqn. \eqref{eqn-simple-kw} then gives~
\begin{align}
  k_x^2
    g(\Vec{k},\omega)
 -
  \bar{\kappa}(\Vec{k},\omega)^2
    g(\Vec{k},\omega)
&=
 -Q'
,
\label{eqn-simple-x-kw-first}
\end{align}
 with $\bar{\kappa}(\Vec{k},\omega)^2 = \omega^2 p(\omega) / s(\Vec{k}) c^2$
 and
~
\begin{align}
 Q'(\Vec{k},\omega)
&=
  Q(\Vec{k},\omega) / s(\Vec{k})
 +
  k_T^2 g(\Vec{k},\omega).
\end{align}

In the converse of the temporal case, 
 where a frequency dependence was inconvenient, 
 here $\bar{\kappa}$ has an inconvenient wavevector dependence;  
 thus it is not a suitable basis for 
 decomposing the wave evolution.
Our choices therefore are to ignore
 the material's spatial response $s(\Vec{k})$, 
 incorporate it into $Q$, 
 or approximate it as a frequency response $s(\omega)$\footnote{For example, 
  if restricted to waves near a wavevector $k_1$
  in a narrow bandwidth $\Delta k \ll k_1$, 
  we can write $k \simeq \omega / c''$, 
  where $c'' = \omega/k_1$, 
  so $s(k)$ can be replaced by $p(\omega)$ --
  i.e. we have converted a spatial dispersion 
  into an approximate temporal dispersion.
 This
  approximation allows the
  spatial dispersion to be added onto the material's temporal dispersion, 
  which is particularly useful
  when propagating waves along a waveguide.}.
If there is no spatial response, 
 then $s(\Vec{r}) \equiv s_0 \delta(\Vec{r})$ and $s(\Vec{k})=s_0$
 so that 
 $\kappa(\omega)^2 = \omega^2 p(\omega) /  s_0 c^2$.
Eqn. \eqref{eqn-simple-t-kw} can be rearranged
 to give an expression for $g(\Vec{k},\omega)$ directly, 
 i.e.
~
\begin{align}
   g
&=
 -
  \frac{1}
       {k_x^2-\kappa^2}
  Q'
&=
 +
  \frac{1}
       {2 \kappa}
  \left[
    \frac{1}{k_x+\kappa}
   -
    \frac{1}{k_x-\kappa}
  \right]
  Q'
,
\label{eqn-Simple-kw-x-g}
\end{align}
Now decompose $g$ into wave components $g^{+}$ and $g^{-}$
 that evolve either forward or backward in time:
 with $g = g^{+} + g^{-}$.
Then split eqn. \eqref{eqn-Simple-kw-x-g}
 into two coupled first order wave equations, 
 i.e.
~
\begin{align}
  g^{\pm}
&=
 \mp
  \frac{1}
       {2 \kappa}
    \frac{1}{k_x\mp\kappa}
  Q'
\\
  \left[
    k_x\mp\kappa
  \right]    
  g^{\pm}
&=
 \mp
  \frac{1}
       {2 \kappa}
  Q'
\label{eqn-Simple-kw-k-pre-gpm}
,
\end{align}
 leading to either wavevector or spatial domain forms
~
\begin{align}
  k_x
  g^{\pm}(\Vec{k},\omega)
&=
 \pm
  \kappa(\omega)
  g^{\pm}(\Vec{k},\omega)
 \mp
  \frac{1}
       {2 \kappa(\omega)}
  Q'(\Vec{k},\omega)
,
\label{eqn-Simple-kw-k-gpm}
\\
  \partial_x
  g^{\pm}(\Vec{r},\omega)
&=
 \pm
  \imath
  \kappa(\omega)
  g^{\pm}(\Vec{r},\omega)
 \mp
  \frac{\imath}
       {2 \kappa(\omega) s_0}
  Q(\Vec{r},\omega)
\nonumber
\\
& \qquad
 \pm 
  \frac{\imath}
       {2 \kappa(\omega)}
  \nabla^2_T 
  \left[
    g^{+}(\Vec{r},\omega) + g^{-}(\Vec{r},\omega)
  \right]
,
\label{eqn-Simple-kw-x-gpm}
\end{align}
where  the different conventions for $\kappa$
 and $\omega$ give rise to the differing signs in the 
 leading (reference) RHS terms if  
 eqns. \eqref{eqn-Simple-kw-t-gpm}
 and \eqref{eqn-Simple-kw-x-gpm} are compared.

This last equation tells us how to evolve $g^{\pm}(\Vec{r},\omega)$ 
 [or if appropriately transformed, $g^{\pm}(\Vec{r},t)$]
 as we propagate along our chosen spatial axis $x$.
Again, 
 by substituting one of these into the other, 
 the original second order equation can be recovered.
Note that any wavevector dependence for $\kappa$
 would have given rise to spatial derivatives on the RHS
 of eqn. \eqref{eqn-Simple-kw-x-gpm}.
In addition to mathematical complications, 
 these would also disrupt
 the otherwise straightforward integration of $g^{\pm}$ along $x$.

Here the directed $g^{\pm}$ 
 evolve according to $\kappa$, 
 but that this reference evolution
 is modified by the additional source terms, 
 either the general term $Q$,
 or the diffraction term dependent on $\nabla_T^2$
 (or $k_T^2$, depending on the chosen domain).
These source terms not only 
 modulate (e.g. drive, amplify, or attenuate)
 the wave equations equally, 
 they also \emph{couple} them together.
\emph{If the source terms are small, }
 i.e. 
\begin{align}
  \left|
    Q/s_0
  \right|
, ~
  \left|
    \nabla_T^2
    \left( g^{+} + g^{-} \right)
  \right|
&\ll
  2
  \left|
    \kappa^2 g^{\pm}
  \right|
,
\end{align}
  then the wave evolves slowly as it propagates.
This ``slow evolution'' permits 
 a unidirectional approximation 
 \cite{Kinsler-2010pra-fchhg,Kinsler-2007josab}
 to be made, 
 setting $g^{-}=0$.
The forward waves $g^{+}(\Vec{r},\omega)$ then follow
~
\begin{align}
  \partial_x
  g^{+}(\Vec{r},\omega)
&=
 +
  \imath
  \kappa(\omega)
  g^{+}(\Vec{r},\omega)
 -
  \frac{\imath}
       {2 \kappa(\omega)}
  \frac{Q(\Vec{r},\omega)}{s_0}
\nonumber
\\
& \qquad
 +
  \frac{\imath}
       {2 \kappa(\omega)}
  \nabla^2_T 
  g^{+}(\Vec{r},\omega)
.
\label{eqn-Simple-kw-x-gp}
\end{align}
Again,
 if $Q$ is an impulse such as a delta function, 
 the slow evolution condition will not hold at that point.
If $\kappa^2 < 0$, 
 then $g$ or $g^{\pm}$ do not oscillate in space,
 but instead are \emph{evanescent}.


%
\subsection{Discussion}\label{S-decompose-discuss}

There is an interesting tension between these two factorizations.
At first sight, 
 the most physical choice of propagation is that in time; 
 however this means at any specific time,
 a full time-history is unavailable.
This means that it is impossible (strictly speaking) to
 know spectral properties
 generally taken for granted, 
 e.g. the wave speed $c(\omega$) or the wavevector $k(\omega)$.
To get frequency-dependent quantities we need either
 (a) a solution containing a complete time history,
 as might be obtained analytically for a some problems,
 or (b) to work in a spatially propagated picture, 
 where a full time history is automatically available.
However, 
 the spatially propagated picture
 differs from 
 our experience of a universe advancing in time.

Further, 
 note that the two factorizations treat diffraction differently -- 
 in the temporally propagated case, 
 having the full spatial profile to hand at each time step 
 means that diffraction can be done exactly, 
 even in a unidirectional model, 
 whereas in the spatially propagated (and unidirectional) case
 a paraxial approximation needs to be applied.
The first order wave equations derived above (and below)
 can be conveniently adapted
 using simple transformations or restrictions\cite{Kinsler-2010pra-fchhg}.
These are typically applied to unidirectional models,
 because whilst they make (e.g.) the representation 
 of the forward wave better behaved, 
 the backward wave becomes more problematic.


%
\section{The time-dependent diffusion equation}\label{S-TDDE}

The time-dependent diffusion equation (TDDE) \cite{Arfken-MMfPhy}
 is a 
 second order wave equation
 with a loss term added; 
 it appears in a variety of contexts in physics, 
 including acoustic waves in plasmas 
 or the interstitial gas filling a porous,
 statistically isotropic, 
 perfectly rigid solid \cite{MorseIngard-Acoustics}.
It has a three-dimensional, 
 inhomogeneous form for the velocity potential $g \equiv g(\vec{r},t)$ of
 \cite{Buckingham-2008jasa}
~
\begin{align}
  \nabla^2
    g
 -
  c^{-2} 
  \partial_t^2
    g
 -
  \eta
  \partial_t
    g
&=
  Q
,
\label{eqn-TDDE-xt}
\end{align}
 where $\nabla^2 = \partial_x^2 +\partial_y^2 + \partial_z^2$.
Here
 { $Q$ is a source term, 
 such as a driving term or some modification to the wave equation.}
Next, 
 $\eta$ is a positive constant that imparts loss,
 and $c$ is the high frequency speed of sound.
In wavevector-frequency space, 
 with $g \equiv g(\vec{k},\omega)$
 and $k^2 = \vec{k} \cdot \vec{k} = k_x^2+k_y^2+k_z^2$, 
 eqn. \eqref{eqn-TDDE-xt} becomes
~
\begin{align}
  c^2
  k^2
    g
 -
  \omega^2
    g
 -
  \imath \omega
  c^2
  \eta
    g
&=
 -
  c^2
  Q
\label{eqn-TDDE-kw}
.
\end{align}

%
\subsection{Temporal propagation, spatial decomposition}\label{S-TDDE-time}

To decompose the TDDE into wave components
 evolving forwards or backwards in space
 first choose to propagate forwards in time
 while utilizing a suitable reference frequency $\Omega(k)$, 
 with $k=|\vec{k}|$.
Eqn. \eqref{eqn-TDDE-kw} for $g(\vec{k},\omega)$ is now written
~
\begin{align}
  \Omega (k)^2
    g
 -
  \omega^2
    g
&=
 -
  c^2 Q
 +
  \imath   
  c^2
  \eta \omega
    g
,
\end{align}
 where $\Omega(k) = c k$, 
 and its reference wave speed is $c_\Omega = \Omega/k = c$.
This is essentially the same as the 
 wave equation eqn. \eqref{eqn-simple-t-kw}, 
 but with $Q' = Q - \imath \eta \omega g$
 replacing $Q$.
As already explained, 
 $\vec{k},t$ domain wave equations can now be given 
 for velocity potentials $g_{+}$ and $g_{-}$
 that evolve forward or backward in space,
~
\begin{align}
  \partial_t
  g_{\pm}
 \mp
  \imath
  \Omega
  g_{\pm}
 \mp
  \frac{\imath c^2}
       {2 \Omega}
  \left[
    Q
   +
    \eta \partial_t
    \left( g_{+} + g_{-} \right)
  \right]
.
\label{eqn-TDDE-kw-t-dgpm}
\end{align}
These directed $g_{\pm}$ 
 evolve according to $\Omega$, 
 but this reference evolution
 is modified by both the general term $Q$
 and the loss term $\eta$.
These source terms not only 
 modulate (e.g. drive, amplify, or attenuate)
 the wave equations equally, 
 they also \emph{couple} them together.
\emph{If the source terms are small, }
 i.e. 
\begin{align}
  \left|
    c^2 Q
  \right|
, ~
  \left|
    c^2 \eta \partial_t
    \left( g_{+} + g_{-} \right)
  \right|
~
&\ll
~ 
  2
  \left|
    \Omega g_{\pm}
  \right|
\label{eqn-TDDE-xw-etaapprox}
,
\end{align}
 the wave evolves slowly as it propagates, 
 so that a unidirectional approximation can be made,
 setting $g_{-}=0$;
 but remain alert to the inconvenient time derivative
 on the left hand side (LHS) of eqn. \eqref{eqn-TDDE-xw-etaapprox}.
The forward wave $g_{+}(\vec{k},t)$ then follows
~
\begin{align}
  \partial_t
  g_{+}
&=
 -
  \imath
  \Omega
  g_{+}
 -
  \frac{\imath c^2}
       {2 \Omega}
  \left[
    Q
   +
    \eta \partial_t
    g_{+}
  \right]
.
\label{eqn-TDDE-xw-t-dgp}
\end{align}

Note that here the RHS side has a time derivative.
Normally it is preferable for all such terms to be on the LHS, 
 so that the RHS directly and unambiguously describes how the wave evolves.
Fortunately, 
 in this unidirectional wave equation,
 the two time derivative terms can be combined
 before 
 re-applying the slow evolution approximations
 to get
~
~
\begin{align}
  \partial_t
  g_{+}
&\simeq
 -
  \imath
   \left(
    \Omega - \frac{\imath c^2 \eta}{2}
   \right)
  g_{+}
 -
  \frac{\imath c^2 }
       {2 \Omega}
    Q
.
\end{align}

%
\subsection{Spatial propagation, temporal decomposition}\label{S-TDDE-space}

To decompose the TDDE into wave components
 evolving forwards or backwards in time,
 first choose to propagate forwards along a spatial axis
 while utilizing a suitable reference wavevector $\kappa(\omega)$.
To do this we need to select a primary propagation direction, 
 here chosen to be along the $x$ axis, 
 so that 
 $k_y$ and $k_z$ are the transverse wavevectors
 (with $k_T^2 = k_y^2 + k_z^2$).

This TDDE contains loss terms dependent on $\eta$, 
 but it can be inadvisable to build loss
 into the reference behaviour
 of spatially propagated waves\cite{Kinsler-2009pra}.
Therefore it is best to
 allow for the possibility that $\eta$ might be 
 retained on either the LHS (in which case $\eta_1=\gamma$ and $\eta_2=0$)
 or the RHS (in which case $\eta_2=\eta$ and $\eta_1=0$).
Eqn. \eqref{eqn-TDDE-kw} for $g(\vec{k},\omega)$ is now written 
~
\begin{align}
  k_x^2
    g
 -
  \kappa(\omega)^2
    g
&=
 -
  Q
-
  k_T^2
    g
 +
  \imath
  \eta_2
  \omega
    g
,
\end{align}
 where $\kappa(\omega)$
 is defined as
~
\begin{align}
  \kappa(\omega)^2
&=
  \frac{\omega^2}{c^2}
  \left[
    1 + \imath \frac{\eta_1}{\omega}
  \right]
,
\end{align}
 so that the lossless ($\eta_1=0$) 
 reference wave speed $c_\kappa = \omega/\kappa$ is constant at $c$.
If complex valued $\kappa$ or $c_\kappa$ are acceptable, 
 then 
~
\begin{align}
  c_\kappa^2
&=
  \frac{c^2}
       {1+\imath \eta_1 / \omega}
\qquad
=
  \frac{\omega c^2}
       {\omega^2 - \eta_1^2}
  \left[
    \omega - \imath \eta
  \right]
.
\end{align}

If a physical justification could be imagined, 
 the spatial decomposition used here would allow the parameters $c, \eta$
 to have a dependence on $\omega$, 
 although the appropriately matching time dependence
 (i.e. convolutions over a temporal history)
 would need to be present in eqn. \eqref{eqn-TDDE-xt} -- 
 as in e.g. eqn. \eqref{eqn-simple-xt}.

After defining $Q' =  Q + k_T^2 g - \imath \eta_2 \omega g$,
 follow the same steps
 as for eqns. \eqref{eqn-Simple-kw-x-g} to \eqref{eqn-Simple-kw-k-pre-gpm}, 
 and decompose $g$ into velocity potentials $g^{+}$ and $g^{-}$
 that evolve forward or backward in time,
 with $g = g^{+}+g^{-}$.
The two coupled first order wave equations
 for $g^{\pm}(\vec{k},t)$ are then
~
\begin{align}
  k_x
  g^{\pm}
&=
 \pm
  \kappa
  g^{\pm}
 \mp
  \frac{1}
       {2 \kappa}
  Q'
\label{eqn-TDDE-kw-z-gpm}
\end{align}
In the  $x, \omega$ domain, 
 eqn. \eqref{eqn-TDDE-kw-z-gpm} can be rewritten
~
\begin{align}
  \partial_x
  g^{\pm}
&=
 \pm
  \imath
  \kappa
  g^{\pm}
 \mp
  \frac{\imath}
       {2 \kappa}
  Q
 \mp
  \frac{\imath k_T^2}
       {2 \kappa}
  \left[
    g^{+} + g^{-}
  \right]
 \pm
  \frac{\imath \eta_2 \omega}
       {2 \kappa}
  \left[
    g^{+} + g^{-}
  \right]
,
\label{eqn-TDDE-kw-x-dgpm}
\end{align}
where  the different conventions for $\kappa$
 and $\omega$ give rise to differing signs in the 
 leading RHS terms 
 (e.g. compare eqns. \eqref{eqn-TDDE-kw-t-dgpm}
 and \eqref{eqn-TDDE-kw-x-dgpm}).

Here the directed $g^{\pm}$ 
 evolve according to $\kappa$, 
 but that this reference evolution
 is modified by the additional source terms, 
 either the general term $Q$,
 the diffraction term dependent on $k_T^2$,
 or the loss term dependent on $\eta_2$.
These source terms not only 
 modulate (e.g. drive, amplify, or attenuate)
 the wave equations equally, 
 they also \emph{couple} them together.
\emph{If the source terms are small, }
 i.e. 
\begin{align}
  \left|
    Q
  \right|
, ~
  \left|
    \eta \omega
    \left( g^{+} + g^{-} \right)
  \right|
, ~
  \left|
    k_T^2
    \left( g^{+} + g^{-} \right)
  \right|
&\ll
  2
  \left|
    \kappa^2 g^{\pm}
  \right|
,
\end{align}
 the wave changes slowly as it propagates in space, 
 so that we can make a unidirectional approximation, 
 setting $g^{-}=0$.
Then the condition containing $k_T^2$
 is making a \emph{paraxial} approximation, 
 which is appropriate where the wave propagates primarily in a narrow beam
 oriented along some particular direction.
The forward wave $g^{+}(x,k_y,k_z,\omega)$ then follows
~
\begin{align}
  \partial_x
  g^{+}
&=
 +
  \imath
  \kappa
  g^{\pm}
 -
  \frac{\imath}
       {2 \kappa}
  Q
 -
  \frac{\imath k_T^2}
       {2 \kappa}
    g^{+}
 -
  \frac{\imath \eta_2 \omega}
       {2 \kappa}
    g^{+}
.
\label{eqn-TDDE-xw-x-dgp}
\end{align}

Here the choice of whether to put
 the loss ($\eta$) dependent term into the reference wavevector $\kappa$
 or not is not necessarily so important
 if the loss is small, 
 since it will not break the unidirectional approximation.
Further, 
 just as for the time-propagated eqn. \eqref{eqn-TDDE-xw-t-dgp},
 again there is a factor of $\omega$ 
 (or if transformed into the time domain,
 a time derivative $\partial_t$)
 applied to the $\eta$ term on the RHS.
Now, 
 however, 
 because we are propagating forward in space, 
 it can be easily calculated using the known $\omega$ (or $t$)
 dependence of $g^{\pm}$.

%
\subsection{Discussion}\label{S-TDDE-discuss}

Both decompositions of the TDDE 
 have the same (lossless) reference wave speed, 
 i.e. the high frequency speed of sound $c = c_\Omega = c_\kappa$.
However, 
 they treat diffraction differently -- 
 in the temporally propagated case, 
 having the full spatial profile to hand at each time step 
 means that diffraction can be done exactly, 
 even in a unidirectional model, 
 whereas in the spatially propagated (and unidirectional) case
 a paraxial approximation needs to be applied.

%
{

%
\section{An elastic rod wave equation}\label{S-ERod}

Another acoustic system to consider is waves
 traveling along an infinite, 
 isotropic and elastic cylindrical rod of radius $R$.
Following Murnaghan's free energy model, 
 Porubov has derived a wave equation governing propagation
 of the solitary waves along such a rod
 \cite{Porubov-ANLSWS}.
There is no impediment 
 in this model against the rod having ``auxetic'' parameters\cite{Kolat-MTW-2010pssb}, 
 e.g. where the Poisson's ratio was negative \cite{Lakes-1987s}. 
This ``elastic rod equation'' (ERE) 
 describes the displacement $g \equiv g(x,t)$ 
 with a second order wave equation of the form
~
\begin{align}
  c^2
  \partial_x^2
    g
 -
  \partial_t^2
    g
 +
  b_1
  \partial_t^2
  \partial_x^2
    g
 -
  b_2
  \partial_x^4
    g
 +
  \chi
  \partial_x^2
    g^2
&=
  Q
.
\label{eqn-ERod-xt}
\end{align}
 where $g$ is  the longitudinal displacement in the rod, 
 and (following \cite{Kolat-MTW-2010pssb}) the coefficients read:
\begin{align}
  c^2 &= \frac{E}{\rho_0}, 
 ~
  \chi = \frac{\beta}{2\rho_0},
 ~
  b_1 = \frac{\nu\left(\nu-1\right) R^2}{2},
 ~
  b_2 = - \frac{\nu E R^2}{2\rho_0},
\\
  \beta 
&= 
  3E 
 +
  l\left(l-2\nu\right)^3
 +
  4m \left(l-2\nu\right)\left(l+\nu\right)
 +
  6n\nu^2
.
\end{align}
Here
 $\beta$ is the nonlinear coefficient, 
 $E$ and $\nu$ are Young's modulus and Poisson's ratio respectively,
 $l$, $m$, $n$ specify Murnaghan's modulus,
 and $\rho_0$ denotes the density.
Poisson's ratio is typically rather small
 (i.e. $|\nu| < 1$), 
 in which case $b_1$ will be negative.
In contrast $b_2$ can cover a wide range of values, 
 especially if auxetic materials are considered, 
 but usually $\nu, E > 0$,
 so that $b_2 <0$.

In eqn. \eqref{eqn-ERod-xt}
{ $Q$ is a source, driving, or other modification to the wave equation.}
In wavevector-frequency space, 
 with $g \equiv g(k,\omega)$,
 eqn. \eqref{eqn-ERod-xt} becomes
~
\begin{align}
  c^2
  k^2
    g
 -
  \omega^2
    g
 -
  b_1
  k^2
  \omega^2
    g
 +
  b_2
  k^4
    g
&=
 -
  Q
 +
  \chi
  k^2
  V_2
\label{eqn-ERod-kw}
,
\end{align}
 where the nonlinear term
 $V_2(k,\omega)$ is derived from $V_2(x,t) = g(x,t)^2$.

%
\subsection{Temporal propagation, spatial decomposition}\label{S-ERod-time}

To decompose the ERE into wave components
 evolving forwards or backwards in space, 
 choose to propagate forwards in time.
This relies on a suitable reference frequency $\Omega(k)$, 
 which results from a careful partitioning
 of the terms in the wave equation \eqref{eqn-ERod-kw}.
The ERE for $g( k,\omega)$ is now written as
~
\begin{align}
  k^2
  \left[
    c^2 + b_2 k^2
  \right]
    g
 -
  \omega^2
  \left[
    1 + b_1 k^2
  \right]
    g
&=
 -
  Q
 +
  \chi
  k^2
  V_2
\label{eqn-ERod-kw-t2initial}
\\
  \Omega (k)^2
    g
 -
  \omega^2
    g
&=
 -Q_k'
,
\end{align}
 where $\Omega(k)$ 
 and new source term $Q_k'$ 
 are 
~
\begin{align}
  \Omega (k)^2
&=
  k^2
  \frac{c^2 + b_2 k^2}
       {1 + b_1 k^2}
~
=
  c^2 k^2 
  \frac{1 + (b_2/c^2) k^2}
       {1 + b_1 k^2}
,
\\
  Q_k'
&=
  \frac{Q}
       {1 + b_1 k^2}
 -
  \frac{\chi k^2 V_2}
       {1 + b_1 k^2}
.
\end{align}

Now follow the same steps
 as for eqns. \eqref{eqn-Simple-kw-t-g} to \eqref{eqn-Simple-kw-w-pre-gpm}, 
 and decompose $g$ into velocity potentials $g_{+}$ and $g_{-}$
 that evolve forward or backward in space, 
 with $g=g_{+}+g_{-}$.
The coupled wave equations
 for $g_{\pm}(k,t)$ are then
~
\begin{align}
  \omega
  g_{\pm}
&=
 \pm
  \Omega
  g_{\pm}
 \pm
  \frac{1}
       {2 \Omega}
  Q_k'
.
\label{eqn-ERod-kw-t-gpm}
\end{align}
Eqn. \eqref{eqn-ERod-kw-t-gpm} can be transformed so as to 
 apply to $g_{\pm}(k,t)$, 
 becoming
~
\begin{align}
  \partial_t
  g_{\pm}
&=
 \mp
  \imath
  \Omega
  g_{\pm}
 \mp
  \frac{\imath}
       {2 \Omega}
  Q_k'
\\
&=
 \mp
  \imath
  \Omega
  g_{\pm}
 \mp
  \frac{\imath}
       {2 \Omega}
  \frac{
    Q
   -
    \chi k^2 V_2}
       {1 + b_1 k^2}
.
\label{eqn-ERod-xw-t-dgpm}
\end{align}
Here the directed displacement waves $g_{\pm}$
 evolve as specified by $\Omega$, 
 but that this reference evolution
 is modified and coupled together 
 by $Q$ and the nonlinear term $\chi V_2$.
Note that $V_2$ needs 
 to be re-expressed in terms of the sum of $g_{\pm}$, 
 a procedure best done in the $x,t$ domain.
\emph{If the source terms are small, }
 i.e. 
\begin{align}
  \left|
    Q 
  \right|
, \quad
  \left|
    \chi k^2 V_2
  \right|
\quad
&\ll 
\quad
  2
  \left|
    \Omega^2 g_{\pm} \left(1 + b_1 k^2\right)
  \right|
,
\end{align}
 then the wave evolves slowly as it propagates, 
 so we can follow the prescription in \cite{Kinsler-2010pra-fchhg}
 and make a unidirectional approximation, 
 setting $g_{-}=0$.
The forward waves $g_{+}(x,t)$ then follow
~
\begin{align}
  \partial_t
  g_{+}
&=
 -
  \imath
  \Omega
  g_{+}
 -
  \frac{\imath}
       {2 \Omega}
  \frac{
    Q
   -
    \chi k^2 V_{2+}}
       {1 + b_1 k^2}
,
\label{eqn-ERod-xw-t-dgp}
\end{align}
 where $V_{2+}$ is just $V_2$ calculated using only $g_{+}$, 
 i.e. with $g_{-} \equiv 0$.
This equation 
 (and indeed the bidirectional eqns. \eqref{eqn-ERod-xw-t-dgpm})
 have a RHS without any frequency (time) dependence, 
 and so give the temporal propagation directly.

%
\subsection{Spatial propagation, temporal decomposition}\label{S-ERod-space}

To decompose the ERE into wave components
 evolving forwards or backwards in time,
 choose to propagate forwards in space.
This relies on a suitable reference wavevector $\kappa(\omega)$, 
 which results from a careful partitioning
 of the terms in the wave equation \eqref{eqn-ERod-kw}.
The ERE for $g(k,\omega)$ is now written as
~
\begin{align}
  \left[
    c^2 - b_1 \omega^2
  \right]
  k^2
    g
 -
  \omega^2
    g
&=
 -
  Q
 +
  \chi k^2 V_2
 -
  b_2 k^4 g
\label{eqn-ERod-kw-x}
\\
  k^2
    g
 -
  \kappa(\omega)^2
    g
&=
 - \frac {\kappa(\omega)^2Q_\omega'}
       {\omega^2}
,
\label{eqn-ERod-kw-x-2owe}
\end{align}
Here $\kappa(\omega)$
 and the new source term $Q_\omega'$
 are defined as
~
\begin{align}
  \kappa(\omega)^2
&=
  \frac{\omega^2} 
       {c^2 - b_1 \omega^2}
,
\\
  Q_\omega'
&=
  Q - \chi k^2 V_2 + b_2 k^4 g
\end{align}

Now following the same steps
 as for eqns. \eqref{eqn-Simple-kw-x-g} to \eqref{eqn-Simple-kw-k-pre-gpm}, 
 we decompose $g$ into velocity potentials $g^{+}$ and $g^{-}$
 that evolve forward or backward in space, 
 with $g = g^{+}+g^{-}$.
The coupled wave equations
 for $g^{\pm}(k,t)$ are 
~
\begin{align}
  k
  g^{\pm}
&=
 \pm
  \kappa
  g^{\pm}
 \mp
  \frac{1}
       {2 \kappa}
  \frac{\kappa^2}{\omega^2}
  Q_\omega'
.
\label{eqn-ERod-kw-z-gpm}
\end{align}
Eqn. \eqref{eqn-ERod-kw-z-gpm} can be transformed so as to 
 apply to $g^{\pm}(x,\omega)$, 
 becoming
~
\begin{align}
  \partial_x
  g^{\pm}
&=
 \pm
  \imath
  \kappa
  g^{\pm}
 \mp
  \frac{\imath \kappa}{2 \omega^2}
  Q_\omega'
\label{eqn-ERod-xw-x-dgpm}
.
\end{align}

Expanding $Q_\omega'$ in eqn. \eqref{eqn-ERod-xw-x-dgpm}, 
 and taking care to transform its $k$ dependence correctly, 
 gives us 
~
\begin{align}
  \partial_x
  g^{\pm}
&=
 \pm
  \imath
  \kappa
  g^{\pm}
 \mp
  \frac{\imath \kappa}
       {2 \omega^2}
  Q 
 \mp
  \frac{\imath \chi \kappa}
       {2 \omega^2}
  \partial_x^2
  V_2
 \mp
  \frac{\imath b_2 \kappa}
       {2 \omega^2}
  \partial_x^4
  \left( g^{+} + g^{-} \right)
\label{eqn-ERod-xw-x-dgpm0}
\end{align}
Again, 
 the directed $g^{\pm}$ 
 evolve according $\kappa$, 
 with this reference evolution
 being modified and coupled together 
 by $Q$,
 the nonlinear term $\chi$,
 and the high-order dispersive term $b_2$.
\emph{If the source terms are small, }
 i.e. 
\begin{align}
  \left|
    Q 
  \right|
, ~
  \left|
    \chi 
    \partial_x^2
    V_2
  \right|
, ~
  \left|
  b_2 
  \partial_x^4
    \left( g^{+} + g^{-} \right)
  \right|
~
&\ll
~
  4
  \left|
    \omega^2 
    g^{\pm}
  \right|
,
\end{align}
 then the wave evolves slowly as it propagates, 
 so we can make a unidirectional approximation, 
 setting $g^{-}=0$.
The presence of the spatial derivatives in the latter two these conditions
 means that more care needs to be taken
 when judging whether or not they are satisfied.
If the conditions can be relied upon to hold, 
 the forward waves $g^{+}(x,\omega)$ then follow
~
\begin{align}
  \partial_x
  g^{+}
&=
 +
  \imath
  \kappa
  g^{+}
 -
  \frac{\imath \kappa}
       {2 \omega^2}
  Q 
 -
  \frac{\imath \chi \kappa }
       {2 \omega^2}
  \partial_x^2
  V_{2+}
 -
  \frac{\imath b_2 \kappa}
       {2 \omega^2}
  \partial_x^4
  g^{+}
.
\label{eqn-ERod-xw-x-dgp0}
\end{align}
This equation 
 has a RHS containing spatial derivatives
 (or, 
  back in the bidirectional eqns. \eqref{eqn-ERod-xw-x-dgpm}, 
  a wavevector dependence), 
 so that the spatial propagation is not given directly.
Consequently,
 for this model, 
 the alternate choice of temporal propagation 
 seems preferable.

%
\subsection{Discussion}\label{S-ERod-discuss}

The nature of the ERE wave equation 
 means that either choice of propagations, 
 based either on $\Omega(k)$ or $\kappa(\omega)$, 
 provide characteristically different evolutions, 
 so that 
 there is no perfect way of connecting the two pictures
 in all circumstances.
In contrast, 
 for the decompositions of the TDDE discussed 
 in section \ref{S-TDDE},
 the reference wave speeds 
 were the same in the lossless case (where $\eta=0$), 
 and similar otherwise.

For a temporally propagated ERE wave,
 the reference wave speed (squared) 
 depends on $\Omega(k)$,
 and is
~
\begin{align}
  c_\Omega^2 
&=
  \Omega^2/k^2
=
  c^2
  \frac{1 + b_2 k^2 / c^2}
       {1 + b_1 k^2}
=
  c^2
  \frac{1 + \sigma b_1 k^2}
       {1 + b_1 k^2}
,
\end{align}
 where 
$ \sigma
=
  b_2/ c^2 b_1 
= -1/(\nu-1) 
= 1/(1-\nu)$.
If $c_\Omega^2 < 0$, 
 as in the case for negative $\sigma$ and large enough $k$,
 waves no longer propagate and instead decay.
At low enough wavevector
 (i.e. where $b_1 k^2, ~\sigma b_1 k^2 \ll 1$), 
 the wave speed is simply $c$.
In contrast, 
 the high-wavevector limit of $c_\Omega^2$
 tends to $\sigma c^2$, 
 which,
 depending on $\nu$ may indicate either propagation or loss, 
 i.e. a pass band or stop band.

The common case where $b_1$ and $\sigma$ are both negative
 is shown on fig. \ref{fig-ERod-ck-compare-b2plus}, 
 along with results for $\sigma>0$.
We we see that the denominator causes an asymptote at $b_1 k^2 = -1$, 
 such that for $\sigma<0$
 a stop band extends from $b_1 k^2 = -1$ upwards; 
 but for $0<\sigma<1$ the stop band has a finite extent, 
 and high-$k$ propagating waves again exist.
For $\sigma>1$ the numerator switches sign first as $k$ increases, 
 and the stop band exist between $\sigma b_1k^1=-1$ and $b_1 k^2 = -1$.
The alternative case where $b_1$ is positive 
 is shown on fig. \ref{fig-ERod-ck-compare-b1plus}, 
 where the low-wavevector limit is simply $c^2$.
In the high-wavevector limit $c_\Omega^2$
 tends to $b_2/b_1$,
 meaning that for $\sigma<0$ (the usual case)
 only low-$k$ waves propagate.

\begin{figure}
\includegraphics[angle=-0,width=0.80\columnwidth]{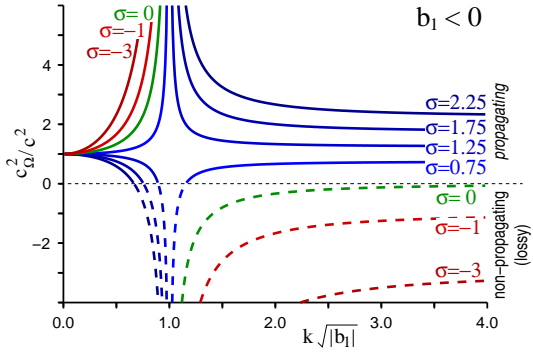}
\caption{
The \emph{temporally propagated} ERE reference wave speed 
 $c_{\Omega}^2$ as a function of wavevector $k$; 
 where both $b_1, \sigma < 0$, 
 for selected $\sigma$.
If $c_\Omega^2 < 0$, 
 the waves decay (dashed lines) instead of propagating.
}
\label{fig-ERod-ck-compare-b2plus}
\end{figure}

\begin{figure}
\includegraphics[angle=-0,width=0.80\columnwidth]{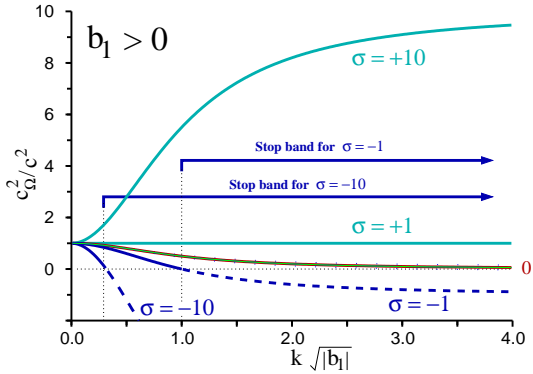}
\caption{
The \emph{temporally propagated} ERE reference wave speed 
 $c_\Omega^2$, 
 in the $b_1>0$ case, 
 for selected $\sigma$.
If $c_\Omega^2 < 0$, 
 the waves do not propagate, 
 but instead decay (dashed lines).
}
\label{fig-ERod-ck-compare-b1plus}
\end{figure}

For the spatially propagated ERE wave, 
 the reference wave speed has a much simpler behaviour; 
 although this simplicity is counteracted by the corresponding wave equations
 (e.g. eqns. \eqref{eqn-ERod-xw-x-dgpm} or \eqref{eqn-ERod-xw-x-dgp0})
 being less straightforward.
The reference wave vector $\kappa^2$ 
 gives a wave speed 
\begin{align}
  c_\kappa^2
&=
  \omega^2/\kappa^2
=
  c^2
  \left(
    1
   -
    b_1 \omega^2 / c^2 
  \right)
.
\end{align}
In the low-frequency limit this is simply $c^2$, 
 but that the other limit is more complicated.
For a high enough frequency 
 (or large enough $b_1$, so that $b_1 \omega^2=c^2$)
 the reference wave speed vanishes.
For higher frequencies, 
 the wave becomes evanescent,
 since $\kappa^2$ is negative and hence $\kappa$ is imaginary.
These are shown on fig. \ref{fig-ERod-cw-compare}.

\begin{figure}
\includegraphics[angle=-0,width=0.80\columnwidth]{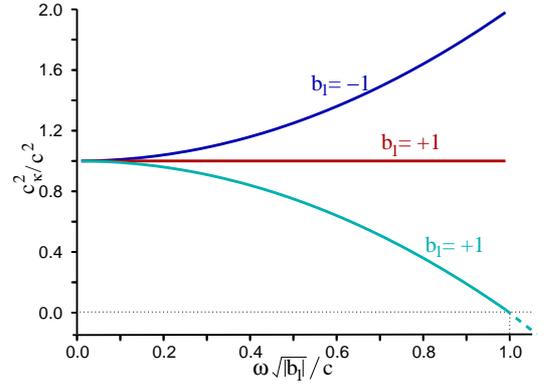}
\caption{
The \emph{spatially propagated} ERE reference wave speed 
 $c_{\kappa}^2$
 as a function of frequency 
 for selected $b_1$.
If $c_\kappa^2 < 0$, 
 the waves are evanescent (dashed line), 
 and do not propagate;
 this occurs for $\omega^2 b_1 / c^2 >1$.
}
\label{fig-ERod-cw-compare}
\end{figure}

Between the two choices of propagation direction
 (i.e. either of time or of space)
 we can see that the partitioning of terms
 leading to the reference frequency or wavevector are not
 the same.
The $b_2$ term only appears in the reference behaviour
 for the choice of temporal propagation, 
 but the $b_1$ term appears in both.
Lastly, 
 despite the differing sign of the $b_1$ term
 (or rather, because of it),
 to first order in $k$ or $\omega$, 
 either reference behaviour 
 gives the same dispersion.

}
%


%
\section{The Stokes' and the van Wijngaarden's equations}\label{S-VWE}

The propagation of an acoustic velocity field
 is commonly described by the Stokes' equation \cite{Stokes-1845tcps}, 
 which can even be 
 be extended to describe propagation of acoustic waves
 in an isothermal, 
 viscous, 
 bubbly liquid \cite{Jordan-F-2006pla}.
This extension was first done for one dimension 
 by van Wijngaarden \cite{Wijngaarden-1972arfm}, 
 hence the ``van Wijngaarden's equation'' (VWE), 
 but has also been generalized to three dimensions \cite{Eringen-1990ijes}.
However, 
 the 3D VWE differs from the common form of the Stokes' equation 
 in that it has spatial operators 
 of $\nabla \nabla \cdot g$ 
 and not $\nabla^2 g$.
Here, 
 to allow for a compact description that 
 includes the 3D Stokes' equation,
 but remains compatible with the 1D VWE, 
 a hybrid 
 Stokes'/ VWE equation (S/VWE) for 
 the velocity potential $g \equiv g(\vec{r},t)$ is introduced.
It is
~
\begin{align}
  \nabla^2
    g
 -
  \frac{1}{c^2}
  \partial_t^2
    g
 +
  \gamma
  \partial_t
  \nabla^2
    g
 +
  \beta^2
  \partial_t^2
  \nabla^2
    g
&=
  Q
.
\label{eqn-VWE-xt}
\end{align}
Here 
{ $Q$ is a source term, 
 such as a driving term or some modification to the wave equation.}
One could even reconcile this equation perfectly
 with the 3D VWE form \cite{Eringen-1990ijes}
 if the differences from the $\nabla^2 \rightarrow \nabla \nabla \cdot$ 
 substitution were incorporated in $Q$, 
 and any side effects properly considered.
The equilibrium bubble radius scales as $\beta$, 
 despite it having dimensions of time. 
With $\beta=0$, 
 this becomes the ordinary Stokes' equation, 
 where the full 3D behaviour of the wave equation is valid.
The parameter
 $\gamma$
 is proportional to the dynamic viscosity of the (bubbly) mixture.
In wavevector-frequency space, 
 with $g \equiv g(\vec{k},\omega)$
 and $k^2 = \vec{k} \cdot \vec{k} = k_x^2+k_y^2+k_z^2$,
 eqn. \eqref{eqn-VWE-xt} becomes
~
\begin{align}
  k^2
    g
 -
  \frac{1}{c^2}
  \omega^2
    g
 -
  \imath 
  \gamma
  k^2
  \omega
    g
 -
  \beta^2
  k^2
  \omega^2
    g
&=
 -
  Q
\label{eqn-VWE-kw}
.
\end{align}

%
\subsection{Temporal propagation, spatial decomposition}\label{S-VWE-time}

\begin{figure}
\includegraphics[angle=-0,width=0.80\columnwidth]{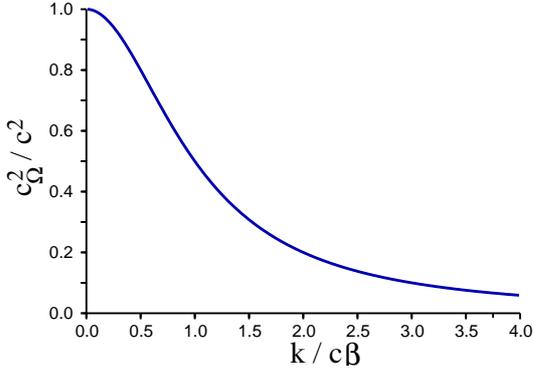}
\caption{
The S/VWE reference wave speed $c_{\Omega}=\Omega/k$
 as a function of scaled wavevector $k/c \beta$.
In the bubble-free Stokes' case, 
 $\beta^2=0$ and the wave speed is a constant at $c_\Omega/c=1$
 for all $k$.
}
\label{fig-FWE-ck-compare}
\end{figure}

To decompose the S/VWE into wave components
 evolving forwards or backwards in space, 
 choose to propagating forwards in time.
This relies on a suitable reference frequency $\Omega(k)$, 
 which results from a careful partitioning
 of the terms in the wave equation \eqref{eqn-VWE-kw}
 to give an $\Omega$ with only a $k$ dependence.
The S/VWE for $g( \vec{k},\omega)$ is now written as
~
\begin{align}
  c^2
  k^2
    g
 -
  \omega^2
  \left[
    1 + c^2 \beta^2 k^2
  \right]
    g
&=
 -
  c^2 Q
 +
  \imath c^2 \gamma \omega k^2 g
\label{eqn-VWE-kw-t2initial}
\\
&=
 - Q'
,
\end{align}
 where $\Omega(k)$ and the new source term $Q'$
 are defined as
~
\begin{align}
  \Omega (k)^2
&=
  \frac{c^2 k^2}
       {1 + c^2 \beta^2 k^2}
,
\\
  Q'
&=
  \left[
    Q/k^2 
   -
    \imath \gamma \omega g
  \right]
  \Omega(k)^2
.
\end{align}
Here 
 $\Omega$ tends to $1/\beta$ in the high-wavevector limit --
 or is constant at $c^2 k^2$ in the Stokes' case where $\beta=0$.
Note that as can be seen on fig. \ref{fig-FWE-ck-compare}
 the reference wave speed $c_\Omega^2 = \Omega^2/k^2$
 in the same high-wavevector limit
 decreases towards zero as $1/k\beta$, 
 and that the low-wavevector limit is simply $c$.
Thus all components of the wavevector spectrum 
 have an ordinary oscillatory (and non-lossy) reference evolution.

Then,
 as for 
 eqns. \eqref{eqn-Simple-kw-t-g} to \eqref{eqn-Simple-kw-w-pre-gpm}, 
 decompose $g$ into velocity potentials $g_{+}$ and $g_{-}$
 that evolve forward or backward in space, 
 with $g = g_{+} + g_{-}$.
The two coupled first order wave equations
 for $g_{\pm}(\vec{k},t)$ are
~
\begin{align}
  \omega
  g_{\pm}
&=
 \pm
  \Omega
  g_{\pm}
 \pm
  \frac{1}
       {2 \Omega}
  Q'
.
\label{eqn-VWE-kw-t-gpm}
\end{align}
Eqn. \eqref{eqn-VWE-kw-t-gpm} can be transformed
 so as to apply to $g_{\pm}(\vec{k},t)$,
 and becomes
~
\begin{align}
  \partial_t
  g_{\pm}
&=
 \mp
  \imath
  \Omega
  g_{\pm}
 \mp
  \frac{\imath}
       {2 \Omega}
  Q'
\\
&=
 \mp
  \imath
  \Omega
  g_{\pm}
 \mp
  \frac{\imath \Omega}
       {2 k^2}
  Q
 \mp
  \frac{\imath \gamma \Omega}{2}
  \partial_t
    \left( g_{+} + g_{-} \right)
.
\label{eqn-VWE-xw-t-dgpm}
\end{align}
Here the directed $g_{\pm}$ 
 evolves as specified by $\Omega$, 
 but that that reference evolution
 is modified and coupled by $Q$ and the dynamic viscosity $\gamma$.
\emph{If the source terms are small, }
 i.e. 
\begin{align}
  \left|
    \frac{\Omega}{k^2} Q
  \right|
, ~~
  \left|
    \gamma \Omega \partial_t
    \left( g_{+} + g_{-} \right)
  \right|
~~
&\ll
~~
  2
  \left|
    \Omega g_{\pm}
  \right|
,
\end{align}
 then the wave evolves slowly as it propagates, 
 so a unidirectional approximation can be made,  
 setting $g_{-}=0$.
The forward waves $g_{+}(\vec{k},t)$ then follow
~
\begin{align}
  \partial_t
  g_{+}
&=
 -
  \imath
  \Omega
  g_{+}
 -
  \frac{\imath \Omega}
       {2 k^2}
  Q
 -
  \frac{\imath \gamma \Omega}{2}
  \partial_t
    g_{+}
.
\label{eqn-VWE-xw-t-dgp}
\end{align}

Notice that a $k$-dependence has appeared on 
 term proportional to $Q$
 in eqn. \eqref{eqn-VWE-xw-t-dgpm} and \eqref{eqn-VWE-xw-t-dgp}, 
 and also that its usual dependence 
 on the reference frequency $\Omega$ has been canceled out.
Although
 an inconvenient RHS time derivative appears, 
 applied to the $\gamma$-dependent term, 
 both time derivatives could be combined 
 to give a directly solvable propagation,
 just
 as for the TDDE model in sec. \ref{S-TDDE}.

%
\subsection{Spatial propagation, temporal decomposition}\label{S-VWE-space}

To decompose the S/VWE into wave components
 evolving forwards or backwards in time, 
 choose to propagate forwards in space.
This uses a suitable reference reference wavevector $\kappa(\omega)$, 
 resulting from a careful partitioning
 of the terms in the wave equation \eqref{eqn-VWE-kw}
 to give an $\kappa$ with only a $\omega$ dependence.
The S/VWE eqn. \eqref{eqn-VWE-kw} for $g(\vec{k},\omega)$
 is now written
~
\begin{align}
  c^2
  \left[
    1 - \imath \gamma \omega - \beta^2 \omega^2
  \right]
  k^2
    g
 -
  \omega^2
    g
&=
 -
  c^2 Q
\label{eqn-VWE-kw-x}
\\
  k^2
    g
 -
  \kappa_0(\omega)^2
    g
&=
 -
  \frac{c^2 \kappa_0(\omega)^2}
       {\omega^2}
  Q
,
\label{eqn-VWE-kw-x-2owe-bare}
\end{align}
 and
 if the propagation axis lies along $x$, 
 this becomes
~
\begin{align}
  k_x^2
    g
 -
  \kappa_0(\omega)^2
    g
&=
 -
  \frac{c^2 \kappa_0(\omega)^2}
       {\omega^2}
  Q
 -
  k_T^2
    g
.
\label{eqn-VWE-kw-x-2owe}
\end{align}
In both eqns. \eqref{eqn-VWE-kw-x-2owe-bare} and \eqref{eqn-VWE-kw-x-2owe}
 the reference wavevector $\kappa_0(\omega)$
 is defined as
~
\begin{align}
  \kappa_0(\omega)^2
&=
  \frac{\omega^2/c^2}
       {1 - \imath \gamma \omega - \beta^2 \omega^2}
.
\end{align}

The S/VWE eqn. \eqref{eqn-VWE-kw-x-2owe}
 contains an imaginary part 
 in its reference wavevector $\kappa_0$, 
 which is not always desirable.
This can be circumvented by
 splitting $\kappa_0^2$ into real and imaginary parts
 using $\kappa_0^2 = \kappa_R^2 + \imath \kappa_I$,
 and moving $\kappa_I$ to the RHS\footnote{Note 
  the imaginary $\kappa_I$ part is \emph{not} squared}:
~
\begin{align}
  \kappa_0(\omega)^2
&=
  \frac{
        \left(1 - \beta^2 \omega^2\right)
        \left(\omega^2 /c^2\right)
       }
       {\left(1 - \beta^2 \omega^2\right)^2 - \gamma^2 \omega^2}
 +
  \frac{\imath \omega \gamma \left(\omega^2/c^2\right)}
       {\left(1 - \beta^2 \omega^2\right)^2 - \gamma^2 \omega^2}
\\
&=
  \kappa_R(\omega)^2
 +
  \imath 
  \kappa_I(\omega)
,
\end{align}

The S/VWE can then be re-expressed as
~
\begin{align}
  k^2
    g
 -
  \kappa_R^2
    g
&=
 -
  \frac{c^2 \kappa_R^2}
       {\omega^2}
    \left(
      1
     +
      \frac{\imath \omega \gamma}
           {1 - \beta^2 \omega^2}
    \right)
    Q
 +
  \imath 
  \kappa_I(\omega)
    g 
,
\label{eqn-VWE-kw-x-2oweR-bare}
\end{align}
although it will also be convenient to merge the $\kappa_I$ term
 into the $Q$ term as follows
~
\begin{align}
  k^2
    g
 -
  \kappa_R^2
    g
&=
 -
  \frac{c^2 \kappa_R^2}
       {\omega^2}
    Q_R
,
\label{eqn-VWE-kw-x-2oweR-QR}
\\
  Q_R
&=
    Q
    \left(
      1
     +
      \frac{\imath \omega \gamma}
           {1 - \beta^2 \omega^2}
    \right)
   -
      \frac{\imath \omega^3 \gamma / c^2}
           {1 - \beta^2 \omega^2}
    g
.  
\label{eqn-VWE-kw-x-define-QR}
\end{align}

If the propagation axis is along $x$, 
 the RHS of the wave equation 
 (still for $g(\vec{k},\omega)$)
 can be written in either of two forms, 
 namely
~
\begin{align}
  k_x^2
    g
 -
  \kappa_R^2(\omega)
    g
&=
 -
  \frac{c^2 \kappa_R(\omega)^2}{\omega^2}
  Q_R
 -
  k_T^2
    g
\label{eqn-VWE-kw-x-2oweR}
\\
&=
 -
  \frac{c^2 \kappa_R(\omega)^2}{\omega^2}
    Q
    \left(
      1
     +
      \frac{\imath \omega \gamma}
           {1 - \beta^2 \omega^2}
    \right)
 +
  \imath 
  \kappa_I
    g
 -
  k_T^2
    g
.
\end{align}

Let us proceed without deciding yet whether $\kappa_0$ and $Q$
 (as in eqn. \eqref{eqn-VWE-kw-x-2owe})
 or $\kappa_R$ and $Q_R$ 
 (as in eqn. \eqref{eqn-VWE-kw-x-2oweR})
 is the best choice.
To achieve this, 
 a subscript-free $\kappa$ and $Q'$ are used
 to stand in for whichever pair of $\{\kappa_0, ~Q\}$
 or $\{\kappa_R, ~Q_R\}$ is convenient, 
 noting that the RHS is given by either of
~
\begin{align}
  \frac{c^2 \kappa_0(\omega)^2}
       {\omega^2}
  Q'
&=
  \frac{c^2 \kappa_0(\omega)^2}
       {\omega^2}
  Q
 + 
  k_T^2 g
,
\\
\textrm{or} \quad
  \frac{c^2 \kappa_R(\omega)^2}
       {\omega^2}
  Q'
&=
  \frac{c^2 \kappa_R(\omega)^2}{\omega^2}
  Q
 -
  \imath
  \kappa_I(\omega)
  g
 + 
  k_T^2 g
\\
&=
  \frac{c^2 \kappa_R(\omega)^2}{\omega^2}
  Q_R
 + 
  k_T^2 g
.
\end{align}
If a physical justification could be imagined, 
 this spatial decomposition would allow the parameters $c, \beta, \gamma$
 to have a dependence on $\omega$, 
 although the appropriately matching time dependence would need
 to be present in eqn. \eqref{eqn-VWE-xt}.

Now following the same steps
 as for eqns. \eqref{eqn-Simple-kw-x-g} to \eqref{eqn-Simple-kw-k-pre-gpm}, 
 decompose $g$ into velocity potentials $g^{+}$ and $g^{-}$
 that evolve forward or backward in time,
 with $g=g^{+}+g^{-}$.
The coupled wave equations
 for $g^{\pm}(\vec{k},\omega)$ are 
~
\begin{align}
  k_x
  g^{\pm}
&=
 \pm
  \kappa
  g^{\pm}
 \mp
  \frac{1}
       {2}
  \frac{c^2 \kappa}{\omega^2}
  Q'
\label{eqn-VWE-kw-z-gpm}
\end{align}
Eqn. \eqref{eqn-VWE-kw-z-gpm} can be transformed 
 so as to apply to $g^{\pm}(x,k_y,k_z,\omega)$, 
 and becomes 
~
\begin{align}
  \partial_x
  g^{\pm}
&=
 \pm
  \imath
  \kappa
  g^{\pm}
 \mp
  \frac{\imath}
       {2}
  \frac{c^2 \kappa}{\omega^2}
  Q'
\label{eqn-VWE-xw-x-dgpm}
.
\end{align}
The two possible choices of $\kappa$ 
 are now addressed separately.

%
\subsubsection{The bare $\kappa_0$ form}\label{S-VWE-space-bareK0}

On the basis of $\kappa_0^2$, 
 the reference wave speed
 in the low-frequency limit is simply $c$, 
 but the other limit is more complicated.
For $\gamma=0$, 
 reference wave speed vanishes at $\omega^2=1/\beta^2$, 
 and above this the wave becomes evanescent,
 since $\kappa_0^2$ is negative and $\kappa_0$ imaginary.
Choosing $\kappa=\kappa_0$ and $Q'=Q+k_T^2 g$, 
 means that eqn. \eqref{eqn-VWE-xw-x-dgpm} 
 for $g^{\pm}(x,k_y,k_z,\omega)$ becomes  
~
\begin{align}
  \partial_x
  g^{\pm}
&=
 \pm
  \imath
  \kappa_0
  g^{\pm}
 \mp
  \frac{\imath c^2 \kappa_0}{2\omega^2}
  Q 
\\
&=
 \pm
  \imath
  \kappa_0
  g^{\pm}
 \mp
  \frac{\imath c^2 \kappa_0}
       {2 \omega^2}
  Q 
 \mp
  \frac{k_T^2}
       {2 \kappa_0}
  \left( g^{+} + g^{-} \right)
.
\label{eqn-VWE-xw-x-dgpm0}
\end{align}
Again, 
 the directed $g^{\pm}$ 
 evolve as specified by $\kappa_0$, 
 but that reference evolution
 is modified and coupled by $Q$ and $k_T^2$.
\emph{If the source terms are small, }
 i.e. 
\begin{align}
  \left|
    \frac{c^2 \kappa_0}{2 \omega^2}
    Q 
  \right|
, ~
  \left|
    \frac{k_T^2}{2 \kappa_0}
    \left( g^{+} + g^{-} \right)
  \right|
~
&\ll
~
  2
  \left|
    \kappa_0 g^{\pm}
  \right|
,
\end{align}
 a unidirectional approximation \cite{Kinsler-2010pra-fchhg} can be made,
 with $g^{-}=0$;
 noting also that the $k_T^2$ condition demands paraxial propagation.
The forward waves $g^{+}(x,k_y,k_z,\omega)$ then follow
~
\begin{align}
  \partial_x
  g^{+}
&=
 +
  \imath
  \kappa_0
  g^{+}
 -
  \frac{\imath c^2 \kappa_0}
       {2 \omega^2}
  Q 
 -
  \frac{k_T^2}
       {2 \kappa_0}
    g^{+}
.
\label{eqn-VWE-xw-x-dgp0}
\end{align}

%
\subsubsection{The dressed $\kappa_R$ form}\label{S-VWE-space-dressedKR}

On the basis of $\kappa_R^2$, 
 the reference wave speed
 in the low-frequency limit is simply $c$, 
 but the other limit is more complicated.
For zero dynamic viscosity (i.e. $\gamma=0$), 
 it matches the bare behaviour, 
 although for finite $\gamma$ the zero-speed transition
 is brought down down to a lower frequency.
Again, 
 above this vanishing reference wave speed 
 the wave becomes evanescent,
 since $\kappa_R^2$ is negative and so $\kappa_R$ imaginary.
With $\kappa=\kappa_R$ and $Q'=Q_R+k_T^2 g$, 
 eqn. \eqref{eqn-VWE-xw-x-dgpm} 
 can be transformed to apply to 
 $g^{\pm}(x,k_y,k_z,\omega)$,
 becoming
~
\begin{align}
  \partial_x
  g^{\pm}
&=
 \pm
  \imath
  \kappa_R
  g^{\pm}
 \mp
  \frac{\imath c^2 \kappa_R}{2\omega^2}
    \left(
      1
     +
      \frac{\imath \omega \gamma}
           {1 - \beta^2 \omega^2}
    \right)
    Q
 \mp
   \frac{\kappa_I}
        {2 \kappa_R}
  \left( g^{+} + g^{-} \right)
\nonumber
\\
& \qquad \qquad \qquad \qquad 
 \mp
  \frac{k_T^2}
       {2 \kappa_R}
  \left( g^{+} + g^{-} \right)
\label{eqn-VWE-xw-x-dgpmR-v1}
\\
&=
 \pm
  \imath
  \kappa_R
  g^{\pm}
 \mp
  \frac{\imath c^2 \kappa_R}{2\omega^2}
    \left(
      1
     +
      \frac{\imath \omega \gamma}
           {1 - \beta^2 \omega^2}
    \right)
    Q
\nonumber
\\
& \qquad 
 \mp
   \frac{\gamma \omega ~ \kappa_R /2}
        {1-\beta^2 \omega^2}
  \left( g^{+} + g^{-} \right)
 \mp
  \frac{k_T^2}
       {2 \kappa_R}
  \left( g^{+} + g^{-} \right)
\label{eqn-VWE-xw-x-dgpmR}
\end{align}
Here the directed $g^{\pm}$ 
 evolve as specified by $\kappa$, 
 but that that reference evolution
 is modified and coupled by $Q$, $\gamma$, and $k_T^2$.
\emph{If the source terms are small, }
 i.e. 
\begin{align}
  \left|
    \frac{c^2}{\omega^2}
    \left(
      1
     +
      \frac{\imath \omega \gamma}
           {1 - \beta^2 \omega^2}
    \right)
    Q 
  \right|
&\ll
  2
  \left|
    g^{\pm}
  \right|
\label{eqn-VWE-slowevolve-Q}
,
\\
  \left|
    \frac{\gamma\omega ~ \left( g^{+} + g^{-} \right) }
         {1 - \beta^2 \omega^2}
  \right|
&\ll
  2
  \left|
    g^{\pm}
  \right|
\label{eqn-VWE-slowevolve-gamma}
\\
\textrm{and} \qquad
  \left|
    \frac{k_T^2}{\kappa_R^2}
    \left( g^{+} + g^{-} \right)
  \right|
&\ll
  2
  \left|
    g^{\pm}
  \right|
\label{eqn-VWE-slowevolve-kT}
,
\end{align}
 then the wave evolves slowly as it propagates.
Note that the constraint on the $Q$ term 
 in eqn. \eqref{eqn-VWE-slowevolve-Q}
 will fail at low frequencies; 
 also that for non-zero $\gamma$, 
 both the $Q$ and the dynamic viscosity (eqn. \eqref{eqn-VWE-slowevolve-gamma})
 conditions will fail
 near $\omega^2 \beta^2 \simeq 1$.
Eqn. \eqref{eqn-VWE-slowevolve-kT} 
 is equivalent
 to demanding paraxial propagation. 
If these conditions, 
 and hence the unidirectional approximation hold,
 it is reasonable to set $g^{-}=0$.
The forward waves $g^{+}(x,k_y,k_z,\omega)$ then follow
~
\begin{align}
  \partial_x
  g^{+}
&=
 +
  \imath
  \kappa_R
  g^{+}
 -
  \frac{\imath c^2 \kappa_R}{2\omega^2}
    \left(
      1
     +
      \frac{\imath \omega \gamma}
           {1 - \beta^2 \omega^2}
    \right)
    Q
\nonumber
\\
& \qquad \qquad \quad
 -
   \frac{\gamma \omega ~ \kappa_R /2}
        {1-\beta^2 \omega^2}
  g^{+}
 -
  \frac{k_T^2}
       {2 \kappa_R}
    g^{+}
.
\label{eqn-VWE-xw-x-dgpR}
\end{align}
Both here in eqn. \eqref{eqn-VWE-xw-x-dgpR}
 and in the bidirectional eqn. \eqref{eqn-VWE-xw-x-dgpmR}
 there is a non-zero $\gamma$-dependent source term.

%
\subsection{Discussion}\label{S-VWE-discuss}

\begin{figure}
\includegraphics[angle=-0,width=0.80\columnwidth]{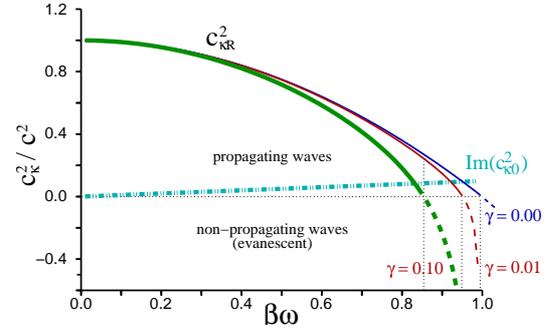}
\caption{
A comparison of the VWE reference wave speed $c_{\kappa R}=\omega/\kappa_R$
 as a function of frequency for different dynamic viscosities $\gamma$
 using the continuous lines;
 note that $c_{\kappa R}=c_{\kappa 0}$ if $\gamma=0$.
Dashed lines indicate non propagating (evanescent) regimes.
The imaginary part of the wave speed squared $c_{\kappa 0}^2$
 is indicated by the dash-double-dotted line, 
for the case where $\gamma=0.1$.
The VWE wave speed for $\gamma=0$ and $\beta=\beta_s$
 is identical to a Stokes' wave speed $c_{\kappa R}$ 
 if $\gamma$ is set equal to $\beta_s$.
}
\label{fig-FWE-cw-compare}
\end{figure}

For this combined S/VWE model, 
 just as for the ERE in section \ref{S-ERod}, 
 the wave equation is such that each choice 
 of propagation direction results in reference evolutions
 that differ both conceptually and in practice.
This is evident from the reference wave speeds,
 which are 
~
\begin{align}
  c_\Omega(k)^2 
&=
  \Omega(k)^2 / k^2 
=
  c^2 / \left( 1 + c^2 \beta^2 k^2 \right)
,
\\
  c_{\kappa 0}(\omega)^2 
&=
  \omega^2 / \kappa_0(\omega)^2 
=
  c^2 \left( 1 - \imath \gamma \omega -  \beta^2 \omega^2 \right)
,
\\
  c_{\kappa R}(\omega)^2 
&=
  \omega^2 / \kappa_R(\omega)^2 
=
  c^2 
  \frac{\left( 1 - \beta^2 \omega^2 \right)^2 - \gamma^2 \omega^2}
       {1 - \beta^2 \omega^2}
.
\end{align}
If $\gamma=0$,
 all describe the same propagation and are in agreement, 
 despite the difference between the high wavevector limit
 where $c_\Omega(k \rightarrow \infty) \rightarrow 0$
 and the transition from propagation to evanescence 
 in $c_\kappa(\omega)$ at $\beta^2 \omega^2=1$.
The behaviour of the frequency-dependent forms $c_\kappa$ are
 shown in fig. \ref{fig-FWE-cw-compare}.

Buckingham \cite{Buckingham-2008jasa} 
 takes issue with the physical properties of analytic solutions
 of the VWE equation, 
 notably its pressure Green's function; 
 this has non-physical high frequency response when $\beta \neq 0$.
Since the $\beta^2$ term in the VWE has a counterpart
 in the ERE's eqn. \eqref{eqn-ERod-xt})
 (i.e. the $b_1$ term), 
 the same remarks could also apply to that model.
However, 
 if the (physically correct) temporal propagation is chosen,
 and either 
 eqns. \eqref{eqn-VWE-xw-t-dgpm}, 
 or 
 eqn. \eqref{eqn-VWE-xw-t-dgp} are 
 \emph{integrated forward in time} --
 the solution has no choice but to be causal.
The only spectral information available at a given time $t$
 depends on the wavevector $k$, 
 \emph{not} the frequency $\omega$.

To understand the frequency spectrum, 
 we must obtain a solution
 to the temporally propagated
 eqns. \eqref{eqn-VWE-xw-t-dgpm} or \eqref{eqn-VWE-xw-t-dgp}, 
 and analyze the temporal history.
That behaviour will be dominated by the reference frequency, 
 which is bounded with $0 \le \Omega(k) < 1/\beta$.
An impulsive $Q$, 
 (e.g. one proportional to a delta function in time), 
 does indeed force onto $g(x,t)$ temporal frequency components above the 
 maximum reference frequency $\Omega_{max} = 1/\beta$, 
 \emph{but the natural behaviour of the wave equation itself will not}.
It is then the spatial profile of the impulse that reveals
 how wave disturbances
 will evolve forward or backward in space, 
 at finite \& \emph{bounded} speeds $c_\Omega(k) \le c$. 
The step discontinuity 
 reported by Buckingham \cite{Buckingham-2008jasa}
 is not a problem with the wave equation, 
 but an image of the impulsive source term driving 
 the velocity potential.

Another issue relates to the case of non-zero $\gamma$, 
 where that loss has been incorporated in the reference behaviour.
In the discussion in subsection III.D
  of Buckingham \cite{Buckingham-2008jasa}, 
 the frequency dependent phase speed diverged as 
 $c_{phase} \simeq 2 c \beta^2 \omega^3 / \gamma$
 for large $\omega$; 
 here this feature can be recovered from the complex $\kappa_0^2(\omega)$, 
 since it is spatial propagation which gives us frequency space properties.
However, 
 if loss is excluded from $\kappa_0$ (and hence $c_{\kappa 0}$), 
 or is zero, 
 or if the dressed $\kappa_R$ ($c_{\kappa R}$) is used,
 then no divergence is seen.
Instead the phase speed is pure imaginary, 
 corresponding to evanescent (and not propagating) waves.
Perhaps most importantly, 
 the wave speed $c_\Omega(k)$ relevant to temporal propagation
 exhibits no such anomaly.

Indeed, 
 given the difficulty of interpreting complex-valued wave speeds, 
 it can be helpful to explicitly exclude loss-like terms
 when evaluating phase and group velocities\cite{Kinsler-2009pra}; 
 i.e. only ever calculate real valued wave velocities.
Directional decompositions are then ideal,
 since they extract easily understood reference behaviors
 from complicated wave equations.

%
\section{Conclusion}\label{S-conclude}

Temporal and spatial propagation schemes
 for some typical acoustic wave equations
 have been compared and contrasted.
The comparison enables a better judgment to be made as to 
 which scheme is more practical in a given circumstance --
 e.g. if reflections are unimportant, 
 a spatially propagated scheme is advantageous due to its 
 more efficient handling of dispersion.
Alternatively, 
 for the ERE, 
 temporal propagation incorporates two parameters ($b_1$, $b_2$)
 into its reference $\Omega(k)$, 
 but spatial propagation incorporates only one ($b_1$) 
 into $\kappa(\omega)$.
This suggests that for the ERE that temporal propagation
 is the natural choice,
 although if material dispersion were present
 the judgment might well become less simple.

Factorization methods mean that 
 such comparisons can be made in a very transparent way --
 structurally similar 
 temporally and spatially propagated wave equations
 can be compared term by term; 
 just as the exact coupled bidirectional wave equations
 can be compared with the approximate unidirectional form
 that results from the single (and physically motivated)
 ``slow evolution'' assumption.
This enables both quantitative and qualitative judgments to be made
 as to the significance of approximations, 
 and/or the effect of any ``source'' terms 
 that perturb or modify the free propagation.
Thus these factorized wave equations have practical advantages 
 for systems that are affected by driving terms, 
 additional material dispersion,
 or nonlinearity -- 
 i.e. effects that make a numerical simulation 
 the most practical way of finding a solution.

The potential for two competing ways of analyzing a situation
 can also add clarity to debate 
 on specific properties of particular acoustic wave equation.
As an example, 
 the apparently non-physical response function for
 the VWE equation \cite{Buckingham-2008jasa} 
 is revealed as an image of the driving impulse, 
 and not of the evolving wave profile.
That profile
 is primarily controlled by $c_\Omega(k)$,
 and not $c_\kappa(\omega)$,
 since to be physically correct the waves must be propagated in time.

In summary, 
 directional decompositions have been used to highlight
 the distinction between 
 the physically accurate choice of temporal propagation 
 and the often more convenient spatial propagation of waves.
Some typical wave equations, 
 containing terms for loss and with high-order derivatives
 have been analyzed, 
 and the handling and consequences of these discussed.


%


\begin{acknowledgments}

I acknowledge support from the
 EPSRC (grant EP/E031463/1)
 and the
 Leverhulme Trust 
 (a 2009 \emph{Embedding of Emerging Disciplines} award).

\end{acknowledgments}

%


\end{document}